\documentclass[aps,preprint,prd,nofootinbib]{revtex4-1}
\usepackage{latexsym,bm,amsmath,amssymb,xfrac,xcolor,tensind}

\usepackage[linktoc=all,hidelinks]{hyperref}

\renewcommand{\thesection}{\arabic{section}}
\renewcommand{\thesubsection}{\arabic{section}.\arabic{subsection}}

\usepackage{titlesec}

\titleformat{\section}
{\normalfont\large\bfseries}{\thesection}{1em}{}
\titleformat{\subsection}
{\normalfont\normalsize\bfseries}{\thesubsection}{1em}{}
\titleformat{\subsubsection}
{\normalfont\normalsize\bfseries}{\thesubsubsection}{1em}{}

\numberwithin{equation}{section}

\tensordelimiter{?}
\tensorformat{}

\def\p{\partial}
\def\m{\mu}
\def\n{\nu}
\def\nab{\nabla}
\def\bnab{\bar{\nabla}}
\def\a{\alpha}
\def\b{\beta}
\def\d{\delta}
\def\D{\Delta}

\def\eps{\epsilon}
\def\f{\phi}

\def\L{\Lambda}
\def\r{\rho}

\def\t{\theta}

\def\vf{\varphi}
\def\nn{\nonumber}

\def\cZ{\mathcal{Z}}
\def\dif{\mathrm{d}}
\def\goesto{\rightarrow}

\def\cL{\mathcal{L}}
\def\cE{\mathcal{E}}

\def\bR{\bar{R}}

\def\bg{\bar{g}}
\def\bnab{\bar{\nabla}}
\def\tT{\widetilde{T}}

\begin{document}
	
		\title{The Kerr-Schild Double Copy in Lifshitz Spacetime}
		
			\author{G{\"o}khan Alka\c{c}}
		\email{gokhanalkac@hacettepe.edu.tr}
		
		\affiliation{Physics Engineering Department, Faculty of Engineering, Hacettepe
			University, 06800, Ankara, Turkey}
		
			\author{Mehmet Kemal G\"{u}m\"{u}\c{s}}
		\email{kemal.gumus@metu.edu.tr}
		
		\affiliation{Department of Physics, Faculty of Arts and Sciences,\\
			Middle East Technical University, 06800, Ankara, Turkey}
	
			\author{Mustafa Tek}
		\email{mustafa.tek@medeniyet.edu.tr}
		
		\affiliation{Department of Physics Engineering, Faculty of Engineering and Natural Sciences, Istanbul Medeniyet University, 34000, Istanbul, Turkey}
\begin{abstract}	
The Kerr-Schild double copy is a map between exact solutions of general relativity and Maxwell's theory, where the nonlinear nature of general relativity is circumvented by considering solutions in the Kerr-Schild form. In this paper, we give a general formulation, where no simplifying assumption about the background metric is made, and show that the gauge theory source is affected by a curvature term that characterizes the deviation of the background spacetime from a constant curvature spacetime. We demonstrate this effect explicitly by studying gravitational solutions with non-zero cosmological constant. We show that, when the background is flat, the constant charge density filling all space in the gauge theory that has been observed in previous works is a consequence of this curvature term. As an example of a solution with a curved background, we study the Lifshitz black hole with two different matter couplings. The curvature of the background, i.e., the Lifshitz spacetime, again yields a constant charge density;  however, unlike the previous examples, it is canceled by the contribution from the matter fields.  For one of the matter couplings, there remains no additional non-localized source term, providing an example for a non-vacuum gravity solution corresponding to a vacuum gauge theory solution in arbitrary dimensions.
\end{abstract}

\maketitle	
\tableofcontents
\newpage
	
\section{Introduction}

The classical double copy is an extension of ideas discovered in the study of scattering amplitudes \cite{Bern:2010ue,Bern:2010yg} to classical solutions of general relativity and gauge theories. The data from the amplitudes suggest a $\text{gravity}=(\text{Yang-Mills})^2$-type relationship where the graviton amplitudes form a double copy of the gluon amplitudes of two Yang-Mills theories that are called single copies. In general, it is possible to relate the classical solutions at a fixed order in perturbation theory \cite{Luna:2016hge,Goldberger:2016iau,Goldberger:2017frp,Goldberger:2017vcg,Goldberger:2017ogt,Shen:2018ebu,Carrillo-Gonzalez:2018pjk,Plefka:2018dpa,Plefka:2019hmz,Goldberger:2019xef,PV:2019uuv,Anastasiou:2014qba,Borsten:2015pla,Anastasiou:2016csv,Cardoso:2016ngt,Borsten:2017jpt,Anastasiou:2017taf,Anastasiou:2018rdx,LopesCardoso:2018xes,Luna:2020adi,Borsten:2020xbt,Borsten:2020zgj,Luna:2017dtq, Kosower:2018adc, Maybee:2019jus, Bautista:2019evw, Bautista:2019tdr, Cheung:2018wkq, Bern:2019crd, Bern:2019nnu, Bern:2020buy, Kalin:2019rwq, Kalin:2020mvi, Almeida:2020mrg, Godazgar:2020zbv, Chacon:2020fmr}; however, for a certain class of spacetimes, a simpler form can be achieved where exact solutions of general relativity are mapped to gauge theory solutions \cite{Monteiro:2014cda, Luna:2015paa, Luna:2016due, Carrillo-Gonzalez:2017iyj, Bahjat-Abbas:2017htu, Berman:2018hwd, Bah:2019sda, CarrilloGonzalez:2019gof, Banerjee:2019saj, Ilderton:2018lsf, Monteiro:2018xev, Luna:2018dpt, Lee:2018gxc, Cho:2019ype, Kim:2019jwm, Alfonsi:2020lub, White:2016jzc, DeSmet:2017rve, Bahjat-Abbas:2020cyb, Elor:2020nqe, Gumus:2020hbb, Keeler:2020rcv, Arkani-Hamed:2019ymq, Huang:2019cja, Alawadhi:2019urr, Moynihan:2019bor, Alawadhi:2020jrv, Easson:2020esh, Casali:2020vuy, Cristofoli:2020hnk}. Recently, a particular version, the so-called Weyl Double Copy \cite{Luna:2018dpt}, was derived through the ideas from twistor theory \cite{White:2020sfn}, implying a much deeper and general relation than previously thought. 

In the pioneering work \cite{Monteiro:2014cda}, the map was obtained by considering solutions of general relativity which can be written in the Kerr-Schild (KS) form with the flat background metric. The fact that the Ricci tensor with mixed indices becomes linear in the perturbation for such solutions provides a natural way to map them to solutions of Maxwell's theory defined on the flat spacetime. A natural extension is to consider spacetimes with non-flat background metrics, which was first studied in \cite{Luna:2015paa}. Later, a more systematic analysis was given in  \cite{Bahjat-Abbas:2017htu} and it was shown that there exist two different ways to realize the double copy structure when the background metric is curved, called Type-A and Type-B double copies. In the Type-A double copy, one maps both the background and the perturbation by using the flat metric as the base. Alternatively, in the Type-B double copy, only the perturbation is mapped by taking the base metric as that of the background spacetime, yielding solutions of Maxwell's theory defined on the curved background. A wide range of examples with constant curvature background was presented in \cite{Carrillo-Gonzalez:2017iyj} where the authors showed the crucial role played by the Killing vectors in the construction. For the stationary solutions, the contraction of the gravity equations with the time-like Killing vector was used, which is essentially checking the $\m0$-components of the trace-reversed equations as done previously. More non-trivial evidence was obtained from the wave solutions where the contraction with the null Killing vector yielded a reasonable single copy.

The linearity of the Ricci tensor in the perturbation, which is the crucial property that makes the whole construction work, holds in the case of a generic curved background spacetime. Motivated by this, in Section \ref{sec:general}, we will give a general formulation of the KS double copy without any simplifying assumption about the background metric. With the assumption that some redundant terms vanish, one obtains Maxwell's equation defined on the curved background where the source term gets a contribution from the curvature of the background, which vanishes for a constant curvature spacetime, in addition to the energy-momentum tensor in the gravity side. In order to see the implications, we will study different solutions of general relativity with a cosmological constant. In Section \ref{sec:max}, solutions with a maximally symmetric background will be examined. When the background is chosen to be of constant curvature, there is no effect on the source. However, choosing a flat background leads to a constant charge density filling all space. While it has been observed before, our formalism explicitly demonstrates that this is due to the deviation of the background from a constant curvature spacetime. In Section \ref{sec:lif}, in order to exhibit the effect of a curved background, we will consider the Lifshitz black hole with two different matter couplings.
\section{General Formulation}\label{sec:general}
In this section, we give a general formulation of the KS double copy in curved spacetime. For that, we will consider classical solutions of cosmological general relativity minimally coupled to matter, which is described by the action
\begin{equation}
	S=\frac{1}{16 \pi G_{d}} \int \text{d}^{d} x \,\sqrt{-g}\,\left[R-2 \Lambda + \cL_m\right],\label{action}
\end{equation}
where $G_{d}$ is the $d-$dimensional Newton's constant, $\L$ is the cosmological constant and $\cL_m$ is the matter part of the Lagrangian density. The field equations arising from the action \eqref{action} are
\begin{equation}
G_{\m\n}+\Lambda\, g_{\m\n}=T_{\m\n}.\label{grav}
\end{equation}
For the KS double copy, one needs the trace-reversed equations with mixed indices
\begin{equation}
	R^\m_{\  \n} -\frac{2\,\Lambda}{d-2}\, \d^\m_{\  \n}=\tT^\m_{\  \n},\label{reversed}
\end{equation}
where the matter contribution is given by
\begin{equation}
	\tT^\m_{\  \n} = T^\m_{\  \n} -\frac{1}{d-2}\,\d^\m_{\  \n}\,T.
\end{equation}
For a metric in the KS form,
\begin{equation}
	g_{\m\n}=\bar{g}_{\m\n}+\phi\, k_\m k_\n,\label{KS}
\end{equation}
where the vector $k_\m$ is null and geodesic with respect to both the background and the full metric as
\begin{equation}
	\bg^{\m\n} k_\m k_\n = g^{\m\n} k_\m k_\n = 0, \qquad \qquad
	 k^\n \bnab_\n k^\m = k^\n \nab_\n k^\m = 0,    
\end{equation}
the Ricci tensor with mixed indices becomes linear in the perturbation as follows \cite{Stephani:2003tm}
\begin{equation}
	R^{\mu}_{\ \nu}=\bar{R}^{\mu}_{\ \nu}-\phi\, k^{\mu} k^{\a} \bar{R}_{\a \nu}+\frac{1}{2}\left[\bar{\nabla}^{\a} \bar{\nabla}^{\mu}\left(\phi \,k_{\a} k_{\nu}\right)+\bar{\nabla}^{\a} \bar{\nabla}_{\nu}\left(\phi\, k^{\mu} k_{\a}\right)-\bar{\nabla}^{2}\left(\phi \,k^{\mu} k_{\nu}\right)\right]\label{ricci}.
\end{equation}
Since the aim is to obtain Maxwell's equations in the background spacetime, we rewrite the Ricci tensor in the KS coordinates \eqref{ricci} by using the gauge field $A_\m \equiv \f\, k_\m$ as
\begin{equation}
	?R^{\m}_{\n}?= {\bR^\mu}_{\ \n}-\frac{1}{2}\left[ \bar{\nabla}_{\a} F^{\a \mu}k_\n+	?E^\m_\n?\right],\label{riccif}
\end{equation}
where
$F_{\m\n}=2\,\bnab_{[\m}A_{\n]}$ is the field strength tensor and 
\begin{equation}
	?E^\m_\n?=?X^{\mu}_\nu?+?Y^{\mu}_\nu?-\bar{R}_{\ \a \b \n}^{\m} A^{\a} k^{\b} +\bar{R}_{\alpha \n} A^{\alpha} k^{\mu},
\end{equation}
with $?X^{\mu}_{\nu}?$ and $?Y^{\mu}_{\nu}?$ given by
\begin{equation}
?X^{\mu}_{\nu}?=-\bar{\nabla}_{\nu}\left[A^{\mu}\left(\bar{\nabla}_{\a} k^{\a}+\frac{k^{\a} \bar{\nabla}_{\a} \phi}{\phi}\right)\right]\,,
\end{equation}
\begin{equation}
?Y^{\mu}_{\nu}? = F^{\a \mu} \bar{\nabla}_{\a} k_{\nu}-\bar{\nabla}_{\a}\left(A^{\a} \bar{\nabla}^{\mu} k_{\nu}-A^{\mu} \bar{\nabla}^{\a} k_{\nu}\right)\,.
\end{equation}
Using this form of the Ricci tensor \eqref{riccif}  in the trace-reversed equations \eqref{reversed} gives,
\begin{equation}
\D^\m_{\  \n} -\frac{1}{2}\left[ \bar{\nabla}_{\a} F^{\a \mu}k_\n+?E^\m_\n?\right] =\widetilde{T}^\m_{\ \n},\label{uncontracted}
\end{equation}
where we introduce the \emph{deviation tensor}
\begin{equation}
\D^\m_{\  \n} =\bR^\m_{\  \n}-\frac{2\,\L}{d-2}\, \d^\m_{\ \n},\label{delta} 
\end{equation}
which vanishes for a constant curvature spacetime if the cosmological constant $\L$ is appropriately chosen, and therefore, characterizes the deviation of the background spacetime from a spacetime with constant curvature (see Appendix for more explanation). 

In order to solve for the field strenght term, we consider the contraction of this equation \eqref{uncontracted}  with a Killing vector $V^\n$ of both the background and the full metric, i.e., 
\begin{equation}
	\nab_{(\m}V_{\n)} = \bnab_{(\m}V_{\n)} = 0,
\end{equation}
which gives the single copy equation as
\begin{equation}
\bar{\nabla}_{\n} F^{\n \mu} + E^\m=J^\m,\label{single}
\end{equation}
where the extra part is
\begin{equation}
	E^\m=\frac{1}{V \cdot k}\,E^\m_{\ \n}\, V^\n, \label{Edef}
\end{equation}
and the gauge theory source is given by
\begin{equation}
J^\m = 2 \left[\D^\m-\tT^\m\right]\label{Jdef},
\end{equation}
with
\begin{equation}
	\D^\m = \frac{1}{V \cdot k}\, \D^\m_{\  \n} V^\n, \qquad \qquad \tT^\m = \frac{1}{V \cdot k}\, \tT^\m_{\  \n} V^\n,\label{deltadef}
\end{equation}
which are the contributions from the background spacetime and the matter part of the Lagrangian respectively.

Contracting the single copy equation \eqref{single} with the Killing vector $V^\m$, one obtains the zeroth copy equation as
\begin{eqnarray}
\bar{\nabla}^2 \phi + \mathcal{Z} +\mathcal{E}  = j,\label{zeroth}
\end{eqnarray}
where
\begin{equation}
	\mathcal{Z} = \frac{V \cdot Z}{V \cdot k},\qquad \qquad \qquad \mathcal{E} = \frac{V \cdot E}{V \cdot k}, \qquad \qquad \qquad j=\frac{V \cdot J}{V \cdot k},\label{defs}
\end{equation}
with vectors $E^\m$ and $J^\m$ given in (\ref{Edef} - \ref{Jdef}) and,
\begin{equation}
Z^\m=\bar{\nabla}_\a k^\m\, \bar{\nabla}^\a \phi + \bar{\nabla}_\a \left[2 \phi \bar{\nabla}^{ [ \a} k^{ \m] } - k^\a \bar{\nabla}^{\m} \phi\right].\label{Zdef}
\end{equation}

 For \emph{any} solution of the gravitational field equations \eqref{grav} that be written in the KS form \eqref{KS}, the gauge field $A_\m=\f \, k_\m$ solves the single copy equation \eqref{single} and the scalar $\f$ solves the zeroth copy equation \eqref{zeroth}. In this paper, we will study black hole solutions in the KS form by using the time-like Killing vector\footnote{In \cite{Carrillo-Gonzalez:2017iyj}, it was shown that the wave-type solutions with maximally symmetric background metrics can be studied by choosing a null Killing vector.} $V^\m=\d^{\m}_{\ 0}$. For the examples that we will consider in this paper, one has
 \begin{equation}
 	V \cdot k = 1, \qquad E^\m=E^{\m}_{\ 0} = 0, \qquad \cE=E^0=0, \qquad \D^\m = \D^\m_{\  0}, \qquad \tT^\m = \tT^\m_{\  0},
 \end{equation}
 and the single copy and the zeroth copy equations becomes Maxwell's and Poisson's equations
 \begin{eqnarray}
 	\bar{\nabla}_{\n} F^{\n \mu}&=&J^\m,\label{maxwell}\nn\\
 	\bar{\nabla}^2 \phi + \mathcal{Z}  &=& j,\label{copies}
 \end{eqnarray}
where the source terms are given by
\begin{equation}
	J^\m =  2 \left[\D^\m-\tT^\m\right],\qquad j=J_0=\bg_{0 \m} J^\m,\label{Jdef2}
\end{equation} 
and  
\begin{equation}
	\cZ=Z_0=\bg_{0 \m} Z^\m,\label{Z}
\end{equation}
with $Z^\m$ given in \eqref{Zdef}. The $\cZ$-term in Poisson's equation vanishes when the background metric is flat and takes a different form depending on the background spacetime. 

The principal result of our analysis is that the deviation of the background metric from a constant curvature spacetime, which is characterized by the deviation tensor defined in \eqref{delta}, affects nontrivially the gauge theory source as described in \eqref{Jdef} for an arbitrary Killing vector and in \eqref{gsource} for the time-like Killing vector. Previously, this has been observed as a constant charge distribution filling all space when the background is taken to be flat. In Section \ref{sec:lif}, we will show that this remains to be true when the background is the Lifshitz spacetime. Therefore, we write the contribution from the background spacetime as
\begin{equation}
\D^\m = \frac{1}{2} \r_{c}\, \d^\m_{\  0},
\end{equation}
where $\r_c$ is the constant charge density. The matter contribution can also be written in the following form
\begin{equation}
\tT^\m = -\frac{1}{2} \r_m v^\m,\label{matcharge}
\end{equation}
where $\r_m$ is the charge density due to the matter in the gravitational theory and $v^\m$ is the velocity of the charge distribution. These lead to the following form of the gauge theory source
\begin{equation}
J^\m = \r_c\, \d^{\m}_{\  0} + \r_m v^\m,\label{gsource}
\end{equation}
which we will use throughout this paper\footnote{As discussed in \cite{Carrillo-Gonzalez:2017iyj}, for black hole solutions, one has localized sources describing a point charge at the origin. Since our main aim is to study the effect of the background spacetime, we will only give the non-localized part of the gauge theory source.}.
 In the next section, we will review some previously studied examples through our general formalism.
\section{Maximally Symmetric Background Spacetime}\label{sec:max}
In this section, we focus on solutions of theories described by the action \eqref{action} with the corresponding field equations \eqref{grav} which can be written in the KS form \eqref{KS} around a maximally symmetric background spacetime. For a non-zero cosmological constant ($\L \neq 0$), the background spacetime can be chosen to be Minkowski or AdS spacetimes. In the former case, the gauge theory copy is defined on Minkowski spacetime and the deviation tensor defined in \eqref{delta} takes the form
\begin{equation}
	\D^{\m}_{\  \n}(\text{Minkowski})=-\frac{2 \L}{d-2} \d^{\m}_{\ \n},\label{delmin}
\end{equation}
since $\bR^\m_{\  \n}=0$, and the constant charge density in the gauge theory source for a timelike Killing vector \eqref{gsource} is determined by the cosmological constant as 
\begin{equation}
\r_c = -\frac{4 \L}{d-2}.\label{rhoc}
\end{equation}
Since the modification to the Poisson's equation given in \eqref{Z} vanishes when the background is Minkowski spacetime, the single copy and the zeroth copy equations become
\begin{eqnarray}
\bar{\nabla}_{\n} F^{\n \mu}&=&J^\m,\nn\\
\bar{\nabla}^2 \phi  &=& j,
\end{eqnarray}
where the general form of the sources is given by
\begin{equation}
J^\m = \r_c\, \d^{\m}_{\  0} + \r_m v^\m, \qquad \qquad j = -(\r_c + \r_m).
\end{equation}
Here, $\r_m$ is the charge density due to the matter fields and the velocity vector $v^\m$ can be read from \eqref{matcharge}. For static solutions, one has a static charge distribution, and therefore, $v^\m = \d^{\m}_{\ 0}$. For stationary solutions, one obtains a rotating charge distribution and the velocity vector takes a form accordingly.

In the latter case, the gauge theory copy is  Maxwell's theory on AdS spacetime and the deviation tensor vanishes
\begin{equation}
	\D^{\m}_{\  \n}(\text{AdS})=0,\label{delads}
\end{equation}
which implies that there is no constant charge density in the gauge theory source ($\r_c=0$). The Poisson's equation is modified due to the curvature of the background as described in \eqref{copies}. In what follows, we will give examples in $d=4$, for which the equations take the following form
\begin{eqnarray}
\bar{\nabla}_{\n} F^{\n \mu}&=&J^\m,\nn\\
\bar{\nabla}^2 \phi  -\frac{1}{6}\bR\, \f &=& j,\label{copyads}
\end{eqnarray}
and the sources are fixed by only the matter contribution as
\begin{equation}
J^\m =  \r_m v^\m, \qquad \qquad j = -\r_m.\label{sourceads}
\end{equation}
In the remainder of this section, we will elaborate on this,  by applying our formalism to some examples that were investigated previously in the literature, with a special focus on the sources, and make a comparison between Minkowski and AdS backgrounds whenever possible.

\subsection{AdS$_4$ Spacetime around Minkowski Background}
As the simplest example, we consider the AdS$_4$ spacetime \cite{Luna:2015paa}, which is a solution when $d=4$, $\cL_m=0$ in \eqref{action}. It can be written in the KS form \eqref{KS} around the Minkowski metric 
\begin{equation}
\bar{g}_{\m\n}\dif x^\m \dif x^\n = -\dif t^2+\dif r^2+r^2\left(\dif \t^2+ \sin^2\t\, \dif \f^2\right),
\end{equation}
where the null vector and the scalar function are given by
\begin{equation}
k_\m \dif x^\m = \dif t+\dif r, \qquad \qquad \f(r)=\frac{\Lambda  r^2}{3}. 
\end{equation}
As a result, the gauge field takes the form
\begin{equation}
A_\m\dif x^\m = \frac{\Lambda  r^2}{3}\, \left(\dif t +\dif x\right),
\end{equation}
and the only non-zero component of the field strength tensor is
\begin{equation}
F_{rt}=\frac{2 \Lambda  r}{3}.
\end{equation}
The effect of the cosmological constant on the sources shows itself as a constant charge density as follows
\begin{equation}
 \r_c = -2\L.
\end{equation}

\subsection{Banados-Teitelboim-Zanelli (BTZ) Black Hole}
An interesting example in three dimensions is the BTZ black hole \cite{Carrillo-Gonzalez:2017iyj}, which is a solution when $d=3$ and $\cL_m=0$ in \eqref{action} for $\L<0$ \cite{Banados:1992wn}. This black hole solution can be obtained by identifyting points of AdS$_3$ spacetime by a discrete subgroup of SO(2,2) \cite{Banados:1992gq} and its gauge theory copy possesses the same characteristics with AdS$_d$ spacetime with $d\geq4$. Its KS form \cite{Kim:1998iw} is given around the Minkowski spacetime in spheroidal coordinates
\begin{equation}
\bar{g}_{\m\n}\dif x^\m \dif x^\n=-\dif t^2+\frac{r^2}{r^2+a^2}\dif r^2+(r^2+a^2)\dif \t^2,\label{min3}
\end{equation}
where $a$ is the rotation parameter. The null vector $k_\m$ is parametrized as
\begin{equation}
k_\m \dif x^\m=\dif t + \frac{r^2}{r^2+a^2} \dif r + a \dif \t,\label{k3}
\end{equation}
and the scalar is given by
\begin{equation}
\phi(r)=1 + 8 G M + \L r^2.
\end{equation}
The corresponding gauge field is given by
\begin{equation}
	A_\m\dif x^\m = \left(1 + 8 G M + \L r^2\right)\left[\dif t + \frac{r^2}{r^2+a^2} \dif r + a \dif \t\right].
\end{equation}
Due to the rotation, there is also a magnetic field and the independent components of the field strenght tensor are 
\begin{equation}
F_{rt}= 2 \L r,\qquad \qquad F_{r\t}= a F_{rt}  = 2 \L a r. 
\end{equation}
The constant charge density corresponding to the BTZ black hole reads
\begin{equation}
 \r_c = -4\L.
\end{equation}
Here, we content ourselves with showing that the constant charge density term appears due to the general property of the deviation tensor \eqref{delmin} and refer the reader to \cite{CarrilloGonzalez:2019gof} for a more detailed discussion.

\subsection{Three-dimensional Rotating Black hole}
Another interesting example from three dimensions is the rotating black hole constructed in \cite{Gumus:2020hbb}. While the background metric and the null vector is the same with our previous example as given in \eqref{min3} and \eqref{k3}, the scalar is given by
\begin{equation}
\phi(r)=-2 M \log r + \L r^2,
\end{equation}
which leads to the gauge field
\begin{equation}
	A_\m \dif x^\m=\left(-2 M \log r + \L r^2\right)\left[\dif t + \frac{r^2}{r^2+a^2} \dif r + a \dif \t\right],
\end{equation}
with the following non-zero components of the field strenght tensor
\begin{equation}
	F_{rt} = 2 \left[-\frac{M}{r}+ \L r\right], \qquad F_{r \t} = a F_{rt} = 2 a \left[-\frac{M}{r}+ \L r\right],
\end{equation}
This solution should be sourced by a space-like fluid\footnote{The static version can be obtained from a free scalar field as the source \cite{Gumus:2020hbb}.} with the following energy momentum tensor
\begin{equation}
T_{\m \n}=(\r+P)u_\m u_\n + P g_{\m \n},
\end{equation}
where
\begin{equation}
P=\frac{M}{r^2}=-\frac{1}{3}\r,\qquad u_\m=\left[\frac{a}{r},0,\frac{a^2+r^2}{r}\right],\qquad u^2 = +1,\label{fluid}
\end{equation}
whose contribution to the trace-reversed Einstein equations is
\begin{equation}
	\tT_{\m\n} = -2 P u_\m u_\n.
\end{equation}
Having a non-zero energy momentum tensor, we get a rotating charge distribution in addition to the usual constant charge density as follows 
\begin{equation}
	\r_c = -2\L, \qquad \qquad \r_m = - \frac{4 M a^2}{r^4}, \qquad \qquad v^\m=\left(1, 0, -\frac{1}{a}\right).
\end{equation} 
\subsection{Schwarzschild-AdS$_4$ Black Hole}
Our next example is the Schwarzschild-AdS$_4$ Black Hole which is a solution with $d=4$ and $\cL_m = 0$ in \eqref{action}. In \cite{Carrillo-Gonzalez:2017iyj}, it was studied around AdS$_4$ spacetime whose metric in global static coordinates reads
\begin{equation}
\bar{g}_{\m\n}\dif x^\m \dif x^\n = -\left[1-\frac{\L r^2}{3}\right]\dif t^2+\left[1-\frac{\L r^2}{3}\right]^{-1} \dif r^2+r^2\left(\dif \t^2+ \sin^2\t\, \dif \f^2\right),\label{ads4}
\end{equation}
and the null vector and the scalar are given by
\begin{equation}
k_\m \dif x^\m = \dif t+\left[1-\frac{\L r^2}{3}\right]^{-1}\dif r, \qquad \qquad \f(r)=\frac{2 M}{r}.\label{kads4} 
\end{equation}
The gauge field
\begin{equation}
A_\m \dif x^\m = \frac{2 M}{r} \left[\dif t+\left[1-\frac{\L r^2}{3}\right]^{-1}\dif r \right],
\end{equation}
has the field strenght tensor with the following non-zero component 
\begin{equation}
F_{rt}=-\frac{2 M}{r^2}.
\end{equation}
We obtain vacuum solutions of \eqref{copyads} since the background is chosen to be of constant-curvature, which implies $\r_c=0$ and there is no contribution from the matter fields ($\r_m=0$). 

The solution can also be written around the Minkowski spacetime
\begin{equation}
\bar{g}_{\m\n}\dif x^\m \dif x^\n = -\dif t^2+\dif r^2+r^2\left(\dif \t^2+ \sin^2\t\, \dif \f^2\right),\label{mink4}
\end{equation}
with the null vector and the scalar defined as
\begin{equation}
k_\m \dif x^\m = dt+dr, \qquad \qquad \f(r)=\frac{2 M}{r}+\frac{\Lambda  r^2}{3}.\label{kmink4}
\end{equation}
The gauge field now becomes
\begin{equation}
A_\m dx^\m = \left[\frac{2 M}{r}+\frac{\Lambda  r^2}{3}\right] \left(\dif t +\dif r\right),
\end{equation}
with the field strenght tensor
\begin{equation}
F_{rt}=-\frac{2 M}{r^2}+\frac{2 \Lambda  r}{3}.
\end{equation}
This time, in the gauge theory source, the only contribution comes from the cosmological constant as
\begin{equation}
\r_c = -2 \Lambda. 
\end{equation}
\subsection{Reissner-Nordstr\"{o}m-AdS$_4$ Black Hole}
In order to see the effect of the matter coupling, we now consider Reissner-Nordstr\"{o}m-AdS$_4$ black hole. The matter part of the action is 
\begin{equation}
	\cL_m = -\frac{1}{4} f_{\m\n} f^{\m \n},
\end{equation}
with contribution to the trace-reversed equations
\begin{equation}
\tilde{T}_{\m \n}=\frac{1}{2} f_{\m\a}f_\n ^{\ \a}-\frac{1}{8} g_{\m\n} f_{\a\b}f^{\a\b}.
\end{equation}
When the metric is written in the KS form around AdS$_4$ spacetime \eqref{ads4} with the null vector given in \eqref{kads4}, the scalar function reads \cite{Carrillo-Gonzalez:2017iyj}
\begin{equation}
\f(r)=\frac{2 M}{r}-\frac{Q^2}{4 r^2},
\end{equation}
where $M$ and $Q$ are the mass and the charge of the black hole respectively. The gauge field becomes
\begin{equation}
	A_\m = \left[\frac{2 M}{r}-\frac{Q^2}{4 r^2}\right]\left[\dif t+\left[1-\frac{\L r^2}{3}\right]^{-1}\dif r \right],
\end{equation}
which leads to the field strength tensor
\begin{equation}
F_{rt}=-\frac{2 M}{r^2}+\frac{Q^2}{2 r^3}.
\end{equation}
While the constant curvature background implies no constant charge density ($\r_c = 0$), the matter field produces the following static charge density
\begin{equation}
\r_m = \frac{Q^2}{2r^4}, \qquad \qquad v^\m = \d^{\m}_{\ 0}.
\end{equation}
One should note that our formalism gives the modification to the Poisson's equation and the source as
\begin{eqnarray}
	\cZ &=& \L\, \frac{Q^2-4Mr^2}{3r^2},\\
	j &=& \frac{Q^2\left(\L r^2 - 3\right)}{6 r^2},
\end{eqnarray}
and one obtains  the standard form given in (\ref{copyads}-\ref{sourceads}) only after simplifications.

When written around the Minkowski spacetime \eqref{mink4} with the null vector \eqref{kmink4}, the scalar function is given by
\begin{equation}
\f(r)=\frac{2 M}{r}-\frac{Q^2}{4 r^2}+\frac{\Lambda  r^2}{3},
\end{equation}
and the gauge field is
\begin{equation}
	A_\m = \left[\frac{2 M}{r}-\frac{Q^2}{4 r^2}+\frac{\Lambda  r^2}{3}\right]\left(\dif t+\dif r\right),
\end{equation}
with the field strength tensor
\begin{equation}
F_{rt}=-\frac{2 M}{r^2}+\frac{Q^2}{2 r^3}+\frac{2 \Lambda  r}{3}.
\end{equation}
In addition to the static charge density $\r_m$ due to the matter part of the Lagrangian, the constant charge density is produced by the non-zero deviation of the Minkowski background \eqref{delmin}, which are given by
\begin{equation}
\r_c = -2\L, \qquad \qquad \r_m = \frac{Q^2}{2 r^4}.
\end{equation}

\section{Lifshitz Black Holes}\label{sec:lif}

So far, we studied metrics that can be written in the KS form around a maximally symmetric background and presented the differences that arise due to the deviation if the Minkowski spacetime from a constant-curvature spacetime, which are a constant charge density in the source and correspondingly, electric and magnetic (if the black hole rotates) fields that linearly increase with the radial coordinate $r$. As an example of a solution with a curved background, in this section, we will consider the Lifshitz black hole in $d$-dimensions, whose metric reads 
\begin{equation}
d s^{2}=L^{2}\left[-r^{2 z} h(r) \dif t^{2}+\frac{\dif r^{2}}{r^{2} h(r)}+r^{2} \sum_{i=1}^{d-2} \dif x_{i}^{2}\right],\label{lifbh}
\end{equation}
where the function $h(r)$ has a single zero at a finite value of $r$ and, $h(r \goesto \infty) = 1$. Asymptotically, the metric takes the form
\begin{equation}
	\left.\dif s^{2}\right|_{r \rightarrow \infty} = L^{2}\left[-r^{2 z} \dif t^{2}+\frac{\dif r^{2}}{r^{2}}+r^{2} \sum_{i=1}^{d-2} \dif x_{i}^{2}\right],\label{lif}
\end{equation}
which is the Lifshitz spacetime. In this form, it is apparent that it describes an asymptotically Lifshitz black hole with a planar horizon. With the following coordinate transformation\footnote{We were informed that the KS form of the Lifshitz black hole was first obtained through this transformation in  \cite{Ayon-Beato:2014wla}.},
\begin{eqnarray}
\dif t \rightarrow \dif  t + \a\, \dif r, \qquad \qquad \a = \frac{h(r)-1}{h(r)} \,r^{-(z+1)},\label{trans}
\end{eqnarray}
one can write the metric in the KS form where the background is the Lifshitz spacetime with the metric
\begin{equation}
\bar{g}_{\mu\nu}\dif x^{\mu} \dif x^{\nu}=L^{2}\left[-r^{2 z} \dif t^{2}+\frac{\dif r^{2}}{r^{2}}+r^{2} \sum_{i=1}^{d-2} \dif x_{i}^{2}\right],\label{lifback}
\end{equation}
The null vector and the scalar are given by
\begin{equation}
k_{\mu} \dif x^{\mu}= \dif t+{\frac{1}{r^{z+1}}} \dif r, \qquad \qquad \phi(r)=L^{2}\,\left[1-h(r)\right]r^{2z}.\label{kandf}
\end{equation}
Note that, for $z=1$, the background metric becomes the AdS spacetime in Poincare coordinates. For $z>1$, the background metric is not maximally symmetric and the deviation tensor will give a non-trivial contribution. The Ricci tensor for the background metric reads
\begin{equation}
	\bR^\m_{\ \n} = \text{diag}\left[-\frac{z\left(z+d-2\right)}{L^2}, -\frac{z^2+d-2}{L^2}, -\frac{z+d-2}{L^2}, -\frac{z+d-2}{L^2}\right],
\end{equation}
which reduces to that of AdS spacetime \eqref{ricads} when $z=1$. The relevant part is still a constant given by
\begin{equation}
	\bR^\m_{\ 0} = -\frac{z\left(z+d-2\right)}{L^2} \d^\m_{\  0},
\end{equation}
which leads to the following background contribution to the gauge theory source
\begin{equation}
	\D^\m\text{(Lifshitz)} = - \left[\frac{z\left(z+d-2\right)}{L^2}+\frac{2\L}{d-2}\right] \d^\m_{\  0},\label{lifdel}
\end{equation}
and, as a result, the following constant charge density
\begin{equation}
	\r_c = - 2 \left[\frac{z\left(z+d-2\right)}{L^2}+\frac{2\L}{d-2}\right].\label{lifrho}
\end{equation}
After this general discussion, we will study two different realizations of the Lifshitz black hole with different matter couplings. 
\subsection{Lifshitz Black Hole from a Massless Scalar and a Gauge Field}

The first solution that we consider is obtained by the following coupling of a massless scalar to a gauge field \cite{Taylor:2008tg}
\begin{equation}
\cL_m=\frac{1}{2} \partial_{\mu} \vf \partial^{\mu} \vf-\frac{1}{4} e^{\lambda \vf} f_{\mu \nu} f^{\mu \nu},
\end{equation}
whose contribution to the trace-reversed equations is
\begin{equation}
\tT_{\m\n} = \frac{1}{2} \partial_{\mu} \vf \partial_{\nu} \vf+\frac{1}{2} e^{\lambda \vf} f_{\mu \a} f_{\nu}^{\ \a}-\frac{1}{4 (d-2)} g_{\mu \nu} e^{\lambda \vf} f_{\a \b} f^{\a \b}.
\end{equation}
The equations for the matter fields are
\begin{eqnarray}
&& \partial_{\mu}\left(\sqrt{-g} e^{\lambda \vf} f^{\mu \nu}\right)= 0, \\
&& \partial_{\mu}\left(\sqrt{-g} \partial^{\mu} \vf\right)-\frac{\lambda}{4} \sqrt{-g} e^{\lambda \vf} f_{\mu \nu} f^{\mu \nu}=0. 
\end{eqnarray}
The Lifshitz black hole is a solution in this theory   with the following metric function \cite{Pang:2009ad}
\begin{equation}
h(r)=1-\frac{r_{+}^{z+d-2}}{r^{z+d-2}}, \qquad \qquad \qquad z\geq 1,\label{met2}
\end{equation}
provided that the matter fields and the cosmological constant are given by
\begin{eqnarray}
f_{r t}&=&q\, e^{-\lambda \vf} r^{z-d+1}, \qquad \qquad e^{\lambda \vf}=r^{\lambda \sqrt{2(z-1) (d-2)}}, \nn \\
\lambda^{2}&=&\frac{2 (d-2)}{z-1},  \qquad \qquad \qquad \, q^{2}=2 L^{2}(z-1)(z+d-2), \nn \\
\Lambda&=&-\frac{(z+d-3)(z+d-2)}{2 L^{2}}.\label{matter2}
\end{eqnarray}
For $z=1$, the matter fields vanish and one obtains the Schwarzschild-AdS black hole  with a planar horizon. By using the  coordinate transformation \eqref{trans}, the metric can be put in the KS form around the Lifshitz background \eqref{lifback} with the null vector and the scalar given in \eqref{kandf}. The explicit form of the scalar for the metric function \eqref{met2} reads
\begin{equation}
 \phi(r)=\frac{L^{2}r_{+}^{z+d-2}}{r^{d-z-2}},
\end{equation}
 The corresponding gauge field is
\begin{equation}
A_\m \dif x^\m = \frac{L^{2}r_{+}^{z+d-2}}{r^{d-z-2}} \left[\dif t+{\frac{1}{r^{z+1}}} \dif r\right],\label{Alif}
\end{equation}
and with the following non-zero component of the field strength tensor
\begin{equation}
F_{r t} = - \frac{(d-z+2) L^2 r_{+}^{z+d-2}}{r^{d-z+1}}.
\end{equation}
Since the matter configuration \eqref{matter2} does not change under the coordinate transformation, it can be directly used in the rest of the calculations. It turns out that the contribution from the deviation tensor and the energy-momentum tensor to the gauge theory source are equal to each other and given by
\begin{equation}
\D^\m = \tT^\m = -\frac{\left(d-3\right)\left(z-1\right)\left(z+d-2\right)}{\left(d-2\right)L^2},
\end{equation}
and therefore, the single copy is
\begin{equation}
\bar{\nabla}_{\n} F^{\n \mu}=0.
\end{equation}
Although we started from a non-vacuum solution, the gauge field given in \eqref{Alif} is vacuum solution of the gauge theory. The modification to the Poisson's equation can again be written in terms of the background Ricci scalar and the KS scalar as
\begin{equation}
\cZ = \frac{z(z-d+2)}{z^2+(d-2)z+\frac{1}{2}(d-1)(d-2)}\bR \,\phi,
\end{equation}
which leads to the following zeroth copy
\begin{equation}
\bar{\nabla}^2 \phi+\frac{z(z-d+2)}{z^2+(d-2)z+\frac{1}{2}(d-1)(d-2)}\bR \,\phi=0.
\end{equation}

\subsection{Lifshitz Black Hole from a Massive Vector and a Gauge Field}
The second solution that we consider is a charged Lifshitz black hole obtained through the following matter coupling \cite{Pang:2009pd}
\begin{equation}
	\cL_m = -\frac{1}{4} f_{\mu \nu} f^{\mu \nu} -\frac{1}{2} m^2 a_\m a^\m -\frac{1}{4} \mathcal{F}_{\mu \nu}\mathcal{F}^{\mu \nu}, 
\end{equation}
where $a_\m$ is a massive vector field with the field strength $f_{\m\n} = 2\, \p_{[\m} a_{\n]}$ and $\mathcal{F}_{\mu \nu}$ is the field strength of the gauge field. The matter field equations are
\begin{eqnarray}
\partial_{\mu}\left(\sqrt{-g}  f^{\mu \nu}\right) &=& m^2 \sqrt{-g}\, a^\n \\
\partial_{\mu}\left(\sqrt{-g}  \mathcal{F}^{\mu \nu}\right)&=&0, 
\end{eqnarray}
and the contribution to the trace-revesed equations is given by
\begin{equation}
\tT_{\m\n}=\frac{1}{2} f_{\mu \a} f_{\nu}^{\ \a}-\frac{1}{4 (d-2)} g_{\mu \nu} f_{\a \b} f^{\a \b} + \frac{1}{2}  m^2 a_{\mu} a_{\nu}
+ \frac{1}{2}  \mathcal{F}_{\mu \a}  \mathcal{F}_{\nu}^{\ \a}-\frac{1}{4 (d-2)} g_{\mu \nu}  \mathcal{F}_{\a \b}  \mathcal{F}^{\a \b} 
\end{equation}
The charged Lifshitz black hole is a solution with the metric function
\begin{equation}
	h(r)=1-\frac{q^2}{2 (d-2)^2 r^z},\label{met1}
\end{equation}
for the matter configuration
\begin{equation}
	a_t=L \sqrt{\frac{2 (z-1)}{z}} h(r) r^z, \qquad \qquad \mathcal{F}_{r t}=q L r^{z-d-1}.\label{matter1}
\end{equation}
The mass of the vector field, the cosmological constant and the Lifshitz exponent should also be fixed as follows
\begin{equation}
	m=\sqrt{\frac{(d-2) z}{L^2}}, \qquad \Lambda=-\frac{(d-3) z+(d-2)^2+z^2}{2 L^2}, \qquad 	z= 2\left(d-2\right).
\end{equation}
The metric can be put in the KS form through the coordinate transformation \eqref{trans} around the Lifshitz background \eqref{lifback}  with the null vector and the scalar given in \eqref{kandf}. The explicit form of the scalar for the metric function \eqref{met2} reads
\begin{equation}
	\phi(r) = \frac{L^2 q^2 r^{3z}}{2 \left(d-2\right)^2}.
\end{equation}
The single copy gauge field and the non-zero component of the field strength tensor are
\begin{eqnarray}
	A_\m \dif x^\m &=& \frac{L^2 q^2 r^{3z}}{2 \left(d-2\right)^2}  \left[\dif t+{\frac{1}{r^{z+1}}} \dif r\right],\\
	F_{r t} &=& \frac{3 z L^2 q^2 r^{3z-1}}{2 \left(d-2\right)^2}. 
\end{eqnarray}
This time, the coordinate transformation \eqref{trans} affects the matter configuration \eqref{matter1} non-trivially, yielding an additional radial component of the massive vector as follows
\begin{equation}
a_r = \a\, a_t.
\end{equation}
The contribution from the deviation tensor and the energy-momentum tensor to the gauge theory source are this time given by
\begin{eqnarray}
	\D^\m &=& -\frac{\left(z-1\right)\left[\left(d-3\right)z+\left(d-2\right)^2\,\right]}{\left(d-2\right)L^2} \d^\m_{\  0},\label{dev}\\
	\tT^\m &=& \D^\m + \frac{q^2}{2 L^2 r^z}\d^\m_{\  0},\label{mat}
\end{eqnarray}
Similar to the previous example, the constant charge density contribution from the deviation tensor \eqref{dev} again disappears, however, this time the contribution from the energy-momentum tensor \eqref{mat} has an additional term, which leads to a non-vacuum solution. The single copy is 
\begin{equation}
	\bar{\nabla}_{\n} F^{\n \mu}=J^\m, \qquad \qquad 	J^\m =  -\frac{q^2}{ L^2 r^z}\d^\m_{\  0}.
\end{equation}
The modification to Poisson's equation in this case can be written as
\begin{equation}
\mathcal{Z} = \frac{z^2}{z^2+(d-2)z+\frac{1}{2}(d-1)(d-2)}\bR \,\phi,
\end{equation}
which yields
\begin{equation}
\bar{\nabla}^2 \phi+\frac{z^2}{z^2+(d-2)z+\frac{1}{2}(d-1)(d-2)}\bR \,\phi=j, \qquad \qquad j = q^2 r^z.
\end{equation}
\section{Summary and Discussions}\label{sec:conc}
In this paper, extending the construction of \cite{Carrillo-Gonzalez:2017iyj}, we gave a formulation of the classical double copy with a generic, curved background spacetime. Apart from obtaining solutions of Maxwell's theory defined on curved backgrounds, our formulation makes the effect of the background spacetime on the gauge theory source much more transparent through the deviation tensor that we defined in \eqref{delta}. For an arbitrary Killing vector of the background and the full metric, the result is given in (\ref{Jdef}-\ref{deltadef}). Choosing a flat background for a solution with a non-zero cosmological constant yields a constant charge density filling all space in the gauge theory due to the general property presented in \eqref{delmin}. The effect disappears when the background is chosen to be a constant curvature spacetime, which can be explained due to the vanishing of the deviation tensor for a suitably chosen cosmological constant \eqref{delads}. Furthermore, we studied two different realizations of the Lifsthiz black hole, whose background is not maximally symmetric. While the contribution to the gauge theory source again turns out to be a constant as described  in (\ref{lifdel}-\ref{lifrho}), it is removed by the matter fields in the gravity side, yielding a vacuum solution in one case.

In the light of our results, there are several directions to pursue. Although the extra part in the single copy equation \eqref{Edef} was shown to vanish for all the examples in the literature, a general proof or, at least, the conditions under which it is true are still lacking. The resolution of this might lead to a better understanding of the classical double copy. The study of wave-type solutions with a curved background, as done in \cite{Carrillo-Gonzalez:2017iyj} for constant curvature backgrounds might also be interesting. In addition to the simplifying assumptions about the background metric, the assumption of the minimal matter coupling can also be released for certain types of theories. We will return to it elsewhere.

\begin{acknowledgments}
	M. K. G. is supported by T\"{U}B\.{I}TAK Grant No 118F091.
\end{acknowledgments}

\appendix
\section{Maximally Symmetric Spacetimes and the Deviation Tensor}
In this appendix, we review some important properties of maximally symmetric spacetimes by following \cite{Natsuume:2014sfa}, which will lead us to the definition of the deviation tensor discussed in the main text. For a maximally symmetric spacetime, the Riemann tensor is given by\footnote{We use barred quantities since, in this work, we consider the possibility of a background metric being that of a maximally symmetric spacetime.}
\begin{equation}
\bR_{\m\a\n\b}= \frac{\eps}{L^2} \left(\bg_{\m\n} \bg_{\a\b}-\bg_{\m\b} \bg_{\n\a}\right)
\end{equation}
where $\eps=+1,0,-1$ correspond to de Sitter (dS), Minkowski and Anti-de Sitter (AdS) spacetimes and $L$ is the dS/AdS radius when $\eps \neq 0$. Taking the trace yield the Ricci tensor and the Ricci scaler as
\begin{eqnarray}
\bR_{\m\n}&=&\eps\, \frac{d-1}{L^2}\,\bg_{\m\n}\label{adsric}\\
\bR&=&\eps\, \frac{d(d-1)}{L^2}
\end{eqnarray}
Using (\ref{adsric}), one can show that the spacetime is a solution of vacuum Einstein equations if the cosmological constant is chosen as
\begin{equation}
\L = \eps \, \frac{(d-1)(d-2)}{2\,L^2}\label{adscosm}
\end{equation}
and, therefore, the Ricci tensor becomes
\begin{equation}
\bR_{\m\n}=\frac{2\,\L}{d-2}\, \bg_{\m\n}.\label{ricads}
\end{equation}
This motivates us to define the deviation tensor from a maximally symmetric spacetime as
\begin{equation}
\D_{\m\n}=\bR_{\m\n}-\frac{2\,\L}{d-2}\, \bg_{\m\n},
\end{equation}
which vanishes for maximally symmetric spacetimes provided that the cosmological constant is given by (\ref{adscosm}). When the background is Minkowski spacetime, one has $\bR_{\m\n}=0$ and,
\begin{equation}
	\D\text{(Minkowski)}_{\m \n} = -\frac{2\,\L}{d-2}\, \bg_{\m\n},
\end{equation}
 which is the origin of the constant charge density in the gauge theory source discussed in the main text.


\end{document}